\documentclass[aps,prl,twocolumn,groupedaddress]{revtex4} 
\usepackage{epsfig,amsopn}
\begin{document}

\title{Dynamic compartmentalization of bacteria:
accurate division in {\it E. coli}.}

\author{Martin Howard}
\altaffiliation[Current address: ]{Instituut-Lorentz, Leiden University,
PO Box 9506, 2300 RA Leiden, The Netherlands}
\affiliation{Department of Physics, Simon Fraser University, Burnaby,
British Columbia, Canada V5A 1S6}
\author{Andrew Rutenberg}
\author{Simon de Vet}
\affiliation{Department of Physics, Dalhousie University, Halifax,
Nova Scotia, Canada B3H 3J5}

\pacs{87.17.Ee, 87.16.Ac, 82.39.Rt}
\begin{abstract} 
Positioning of the midcell division plane within the bacterium
{\it E. coli} is controlled by the
{\it min} system of proteins: MinC, MinD and MinE.
These proteins {\it coherently} oscillate from end to end of
the bacterium. We present a reaction--diffusion model describing
the diffusion of {\em min} proteins along the bacterium  and their 
transfer between the cytoplasmic membrane and
cytoplasm. Our model spontaneously generates protein oscillations in
good agreement with experiments. We explore the oscillation 
stability, frequency and wavelength as a function of
protein concentration and bacterial length.
\end{abstract}  
\maketitle

\vspace*{-3mm}

The subcellular spatial and temporal 
organization of bacterial proteins is largely unknown. 
Already, the spatial distribution of proteins on the cytoplasmic
membrane of bacteria are known to be
important for chemotaxis \cite{Maki2000} and for DNA replication
\cite{Lemon98}.  Improving our understanding of how this supra--molecular 
organization of proteins affects bacterial function
represents a considerable experimental and theoretical challenge. 
In contrast to nucleated eukaryotic cells, no large
organelles are present in the bacterial interior (cytoplasm) and no active
transport mechanisms such as molecular motors are known to function there. 
However, recent video 
microscopy of fluorescently labeled proteins involved in the regulation of
{\em E. coli} division have uncovered coherent and stable spatial and
temporal {\em oscillations} in three proteins: MinC, MinD,
and MinE \cite{Raskin,Raskin1,Hu,Rowland,Fu,Hale}. The proteins
oscillate from end to end of the bacterium, and move between the cytoplasmic 
membrane and the cytoplasm. These {\em min}
proteins select the site for the next bacterial division
\cite{deBoer1,Rothfield1}. Despite a wealth of phenomenological detail,  
no quantitative models have been developed of how the {\em min} proteins 
organize into oscillating structures. 
Understanding the self--organized 
patterns involved in bacterial division processes can give us insight into
how a bacterium can dynamically compartmentalize itself. 

We focus on {\it E. coli}, a commonly--studied rod shaped bacterium,
approximately $2-6~\mu m$ in length and around $1-1.5 \mu m$ in diameter.
Each {\it E. coli} divides roughly every hour, depending on the conditions
 --- first replicating its DNA then dividing in half to form
two viable daughter cells.  The MinCDE oscillations 
are known to persist even when protein synthesis is 
suppressed \cite{Raskin}, and DNA replication and septation occur even 
without the {\em min} proteins. Hence the {\it min} system can be
studied independently of the other division processes. Efficient
division requires many processes, 
including DNA replication, MinCDE oscillations, and
the actual septation process. Septation initiates with a contractile polymeric
``Z--ring'' of a tubulin--homologue FtsZ that forms just underneath the 
cytoplasmic membrane.  The FtsZ septation rings largely
avoid guillotining the DNA-containing nucleoids independently of 
the {\em min} system \cite{Yu99}. This ``nucleoid occlusion''
serves as a complementary control mechanism for accurate cell division. 
The role of the {\em min} system appears to be to restrict the 
Z--ring to midcell. This
reduces the production of inviable nucleoid--free minicells which
occur when the cell divides too close to the cell poles.  If the 
{\em min} system is genetically knocked out, 40\% of divisions lead
to inviable minicells \cite{deBoer1} -- a sizeable drain on 
bacterial resources. 

The study of deletion mutants has made
the phenomenological roles of the individual {\em min} proteins clear. MinC
associated to the cytoplasmic membrane locally inhibits assembly 
of the contractile Z--ring, but remains cytoplasmic and largely
inactive in the absence of MinD \cite{Hu}.  MinD binds MinC and 
recruits it to the cytoplasmic membrane \cite{Hu,Huang}.  MinE drives
MinD away from the bacterial midplane, and hence allows a contractile
ring to form only there.  Without MinE, the membrane--bound MinC/MinD 
block Z--ring formation everywhere, inhibiting division, and resulting in the
formation of long filamentous cells \cite{Raskin1,Rowland}.
Without MinC, Z--ring formation cannot be inhibited anywhere and inviable
minicells are produced. Without MinD, neither MinC nor MinE are
recruited to the cytoplasmic membrane and so have reduced effect.

With normal levels of MinC, D, and E, a remarkable oscillatory
dynamics is seen \cite{Raskin,Raskin1,Hu,Rowland,Fu,Hale}. 
First the MinC/MinD accumulate at one end of the bacterium on 
the cytoplasmic membrane. Then MinE forms a
band at midcell which sweeps towards the cell pole
occupied by the MinC/MinD, ejecting the MinC/MinD into the cytoplasm 
as it goes.  The ejected MinC/MinD then rebinds at the other end of the 
bacterium.  When the MinE band reaches the cell pole, it disassociates and 
reforms at midcell.  The entire process then repeats towards the opposite 
cell pole.  The oscillation period is approximately $1-2$ minutes, so many 
oscillations occur between each bacterial division.  The dynamics 
minimizes the MinC/MinD concentration at midcell, 
thereby allowing the Z--ring and the subsequent division septum to form 
there.

Until recently \cite{Hale}, qualitative models of the {\it min} system
involved unidentified midcell topological markers (see, e.g.,
\cite{Rothfield}). This letter puts forward the first quantitative 
self--organized model that describes much of the intricate
phenomenology of accurate division site placement in {\it E. coli},
and does so using {\em only} the diffusive motion and interactions of
the {\em min} proteins. The essence of our approach is to describe the
MinCDE dynamics by a set of coupled reaction--diffusion equations.
Experimental results indicate that the oscillatory
protein dynamics is unaffected if new protein synthesis is blocked
\cite{Raskin}. Accordingly we employ a model that conserves the total
number of each protein type. Strikingly, this model possesses a
linear Turing--like (Hopf) instability \cite{Cross,Murray} despite the
absence of mechanisms such as internal reactant production or external
feed that have normally been required to model Turing patterns
\cite{Meinhardt}. [Of course energy input in the form of ATP is 
required to sustain the oscillations within a bacterium.] 
As we will see, the resulting protein oscillations
mark the midcell with a minimum of the time--averaged concentration
of MinC/MinD and with a corresponding maximum of MinE. 

Our starting point is a set of four coupled 
reaction--diffusion equations describing, respectively, the
densities of MinD on the cytoplasmic membrane ($\rho_d$), MinD in the
cytoplasm ($\rho_D$), MinE on the cytoplasmic membrane ($\rho_e$), and
MinE in the cytoplasm ($\rho_E$): 
\begin{eqnarray}
{\partial\rho_{D}\over\partial t}& =& D_{D}{\partial^2 \rho_{D}\over
\partial x^2} -{\sigma_{1}  
\rho_{D}\over 1+\sigma_{1}'\rho_{e}}+\sigma_{2} \rho_{e} \rho_{d}
\label{rdD} \\ 
{\partial\rho_{d}\over\partial t}& = & \hspace{1.62cm} \ {\sigma_{1} 
\rho_{D}\over 1+\sigma_1'\rho_{e}}-\sigma_{2} \rho_{e} \rho_{d}
\label{rdd} \\ 
{\partial\rho_{E}\over\partial t}&=&D_{E}{\partial^2\rho_{E}\over
\partial x^2} -\sigma_{3} 
\rho_{D}\rho_{E}+{\sigma_{4}\rho_{e}\over 1+\sigma_{4}'\rho_{D}}
\label{rdE} \\ 
{\partial\rho_{e}\over\partial t}&= &\ \hspace{1.62cm} \ \sigma_{3} 
\rho_{D}\rho_{E}-{\sigma_{4}\rho_{e}\over 1+\sigma_{4}'\rho_{D}} .
\label{rde} 
\end{eqnarray}
Following the observation in Refs.~\cite{Raskin1,Hu} 
that the MinC dynamics simply follows that of the MinD, we do not
model the MinC field explicitly.  We consider the
variation of density along the long bacterial axis, tracking the local
rates of change of the densities stemming from diffusion and from
transfer between the cytoplasmic membrane and the cytoplasm. Zero flux
``closed'' boundary conditions are imposed at both ends of the bacterium. 
The total amount of MinD and MinE, obtained by integrating
$\rho_d+\rho_D$ and $\rho_e+\rho_E$ over the length of the bacterium,
is explicitly conserved by our dynamics.

By reducing the {\em min} protein dynamics to a 
set of deterministic $1d$ rate equations we neglect fluctuation
effects.  Given that the number of {\it min}
molecules in each cell is rather small (around 3000 for MinD
\cite{deBoer} and 170 for MinE \cite{Zhao})
these fluctuations could be important.  While 
some fluctuation effects are evident 
experimentally, such as an occasional mid--cycle reversal of the direction of
MinE band propagation \cite{Hale}, 
on the whole bacterial oscillations appear to be amazingly regular
\cite{Hu}.  Our continuum coarse--grained approach captures the essence  
of the protein dynamics and explains the self--organized
aspects of the MinCDE oscillations.

In the first reaction terms in Eqs.~(\ref{rdD}) and (\ref{rdd}),
$\sigma_1$  describes the spontaneous association of MinD to the
cytoplasmic membrane \cite{Rowland}. MinD is required to recruit 
MinE to the cytoplasmic membrane, but it is an open question whether it is
cytoplasmic MinD or membrane--bound MinD that is primarily active.  
A cytoplasmic interaction between MinD and MinE has been observed in
Ref.~\cite{Huang}, and we are currently only able to obtain the 
MinCDE oscillations by allowing cytoplasmic MinD to recruit cytoplasmic
MinE to the membrane, via $\sigma_3$ in
Eqs.~(\ref{rdE}) and (\ref{rde}). Once on the membrane, 
MinE drives MinD into the cytoplasm. We represent
this with $\sigma_2$ in the second reaction terms in
Eqs.~(\ref{rdD}) and (\ref{rdd}). Finally, MinE will spontaneously
disassociate from the membrane, corresponding to $\sigma_4$ in the
second reaction terms in Eqs.~(\ref{rdE}) and (\ref{rde}).  
We have not included spontaneous MinD disassociation or spontaneous
MinE association terms, since experimentally MinE dominates the MinD 
disassociation and MinD dominates the MinE association.  

Many other reaction terms are possible, however we only include 
the simplest possible ``renormalizations'' of the basic recruitment 
and release terms, $\sigma_1'$ and $\sigma_4'$.  
Effectively $\sigma_1'$ 
corresponds to membrane--bound MinE suppressing the recruitment 
of MinD from the cytoplasm, and $\sigma_4'$
to cytoplasmic MinD suppressing the release of membrane--bound
MinE.
We have also set the diffusion constants for the 
membrane--bound MinD and MinE to zero.  Our results are not
qualitatively changed by using nonzero values, provided the membrane
diffusion constants remain much less than their bulk counterparts.

For our simulations we discretized space and time with spacings of $dx=8\times
10^{-3} \mu m$ and $dt=1\times 10^{-5} s$. We have checked that our
results are unchanged with smaller $dx$ and $dt$. Densities are measured
in molecules per micron, and unless otherwise stated we use average
densities of $1500~\mu m^{-1}$ for
MinD \cite{deBoer} and $85\mu m^{-1}$ for MinE \cite{Zhao}. The numerical
values of our other parameters have not been experimentally determined
for the {\em min} proteins.  We choose cytoplasmic diffusion constants
slightly less than the value $2.5~\mu m^2 s^{-1}$ directly measured 
for a maltose binding protein \cite{Elowitz} within the 
{\em E. coli} cytoplasm.  Unless otherwise mentioned 
we use a length of $2 \mu m$ and the following values for the parameters in
Eqs.~(\ref{rdD})-(\ref{rde}): $D_D=0.28~\mu m^2/s$, $D_E=0.6~\mu m^2/s$, 
$\sigma_1=20~s^{-1}$, $\sigma_1'=0.028~\mu m$, $\sigma_2=0.0063~\mu m/s$,
$\sigma_3=0.04~\mu m/s$, $\sigma_4=0.8~s^{-1}$, and $\sigma_4'=0.027~\mu m$. 

We have analyzed the linear stability of 
Eqs.~(\ref{rdD})-(\ref{rde}) \cite{Murray}.
Testing solutions of the form $e^{\lambda t + iqx}$ 
with the above parameter values,
we find a complex $\lambda(q)$ with a positive real part that is 
maximized for $q\approx 1.5 \mu m^{-1}$, where 
$\lambda_{\rm max}=0.010\pm 0.043i$.  This indicates the 
presence of a maximally linearly unstable oscillating mode with a
wavelength of $4.2~\mu m$ and period of $145~s$. This finding is confirmed
by a direct numerical stability analysis of our model (not shown). 
The physical origin of this instability lies in the disparity between the
membrane and cytoplasmic diffusion rates, and also in the slower rate at
which MinE disassociates from the membrane. This ensures that the MinE
dynamics lags that of the MinD, setting up the oscillating patterns.
The existence of the linear instability in Eqs.~(\ref{rdD})-(\ref{rde})
is crucial, since it means that the oscillating pattern will
spontaneously generate itself from a variety of initial conditions ---
including nearly homogeneous ones. 
In our simulations we used random initial conditions, although
identical patterns were also observed with asymmetric initial
distributions of MinD and MinE. 
The eventual oscillating state is stabilized by 
the nonlinearities in Eqs.~(\ref{rdD})-(\ref{rde}). 
At the midcell, this oscillating pattern has a minimum of the 
time--averaged MinD concentration --- an 
essential feature of division regulation --- and a 
maximum of the time--averaged MinE concentration.

\begin{figure}[h]
\epsfxsize=3.5in
\epsfbox{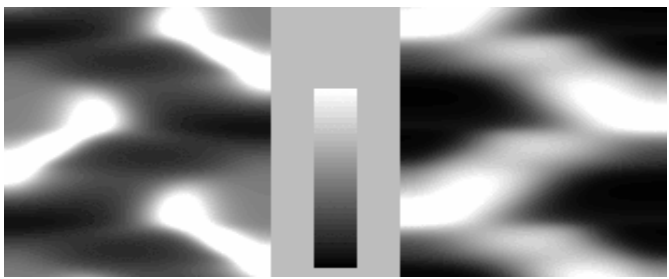}
\caption{Space--time plots of the total MinD (left) and MinE
(right) densities.  The greyscale runs from 0.0 to 2.0 times the
average density of MinD or MinE, respectively.
The MinD depletion from midcell and the
MinE enhancement at midcell are immediately evident. Time increases
from top to bottom, and the pattern repeats indefinitely as time
increases. The greyscale reference bar spans 100 sec.
The horizontal scale spans the bacterial length ($2~\mu m$).}
\label{spacetime}
\end{figure}

\begin{figure}[h]
\epsfxsize=3.5in
\epsfbox{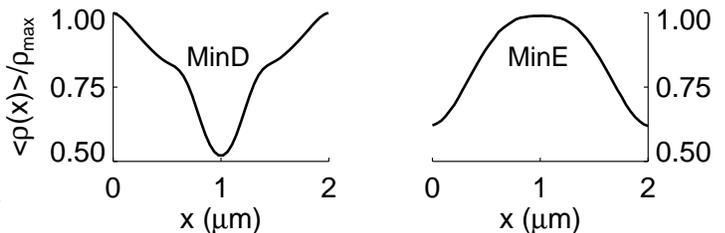}
\caption{The time-average MinD (left) and
MinE (right) densities, $\langle \rho(x) \rangle/\rho_{max}$,
relative to their respective time-average maxima,
as a function of position $x$ (in $\mu m$) along the bacterium.}
\label{average}
\end{figure}

Space--time plots of the MinD
and MinE concentrations for a cell length of $2~\mu m$ are shown in 
Fig.~\ref{spacetime}.  
In excellent agreement with the experimental results, the MinE
spontaneously forms a single band at midcell which then sweeps towards a
cell pole, displacing the MinD, which then reforms at the opposite
pole. Once the MinE band reaches the cell pole it disappears into the 
cytoplasm, only to
reform at midcell where the process repeats but in the other half of
the cell. These patterns are stable over at least $10^9$ iterations 
($10^4~s$) --- long enough for the {\it min} system to regulate 
cell division throughout the division cycle of the cell. 
In Fig.~\ref{average}, we 
plot the time--averaged MinD and MinE densities
as a function of position. MinD shows a pronounced dip in 
concentration close to midcell, which allows for the removal of
division inhibition at midcell. This is in qualitative agreement with the
experimental data of Ref.~\cite{Hale}. MinE peaks 
at midcell, with a minimum at the cell extremities.

We also investigated longer filamentous bacteria and found a
multiple MinE band structure (not shown).  Multiple MinE bands 
always combined into a single MinE band 
in cell lengths shorter than the natural
wavelength indicated by linear stability analysis.

\begin{figure}[h]
\epsfxsize=3.4in
\epsfbox{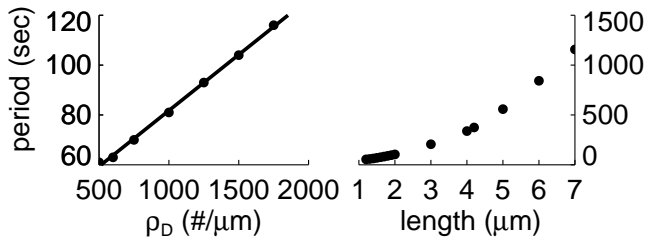}
\caption{Left: Plot of the period of oscillation (in seconds) against MinD
density (in $\mu m^{-1}$), at fixed average MinE concentration of 
$85~\mu m^{-1}$. The solid line is a linear best fit. Right: Plot of
oscillation period against cell length, for fixed MinD and MinE concentrations.
Below bacterial lengths of $1.2 \mu m$ oscillation is not observed.} 
\label{period_density}
\end{figure}

The oscillation period 
as a function of the average MinD concentration is shown in
Fig.~\ref{period_density} (left).
We find a linear relationship indicated by the best--fit line, where
the period approximately doubles as the MinD concentration is 
quadrupled.  A linear relationship has also been suggested
experimentally~\cite{Raskin}.  The period of oscillation as a
function of cell 
length is shown in Fig.~\ref{period_density} (right).  Below lengths of 
$1.2~\mu m$ the bacterium does not sustain oscillating
patterns. For lengths above this minimum, the
oscillation patterns are stable and the period increases with
length --- as observed experimentally \cite{Fu}. The periods measured
from our numerics 
for cell lengths of $2~\mu m$ are around $100~s$, in
good agreement with experiments, where periods from $30-120~s$ have
been found \cite{Raskin}. A single MinE band
state is stable over a wide range of lengths for a given density of
{\it min} proteins. This provides strong evidence that the {\it min}
system is capable of regulating accurate cell division over normally
occurring cell lengths as the cell grows between division events.
At longer lengths of around $6~\mu m$ we observe long--lived metastable
states with two MinE bands. These multiple bands can survive for a
thousand seconds or more before decaying into a single band. At still
longer lengths the two band state appears stable; 
this occurs around $8.4~\mu m$ --- twice the dominant wavelength given
by the linear stability analysis. This explains why the characteristic
wavelength of linear stability analysis is rather longer
than a normal {\em E. Coli} bacterium -- if the length scale were smaller
then multiple MinE bands might occur in bacteria of normal lengths
and proper division regulation would be inhibited.

\begin{figure}[h]
\epsfxsize=3.4in
\epsfbox{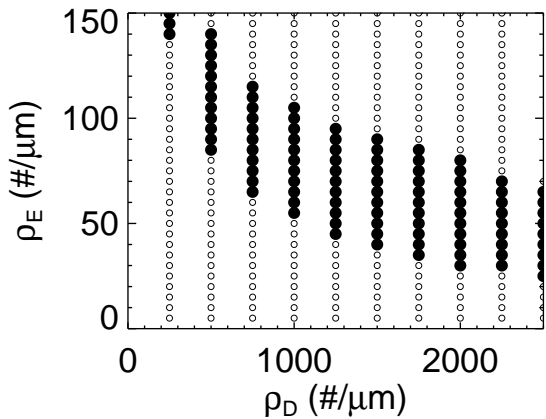}
\vspace*{-5mm}
\caption{Filled circles indicate regions of linear instability vs. 
the density of MinD and MinE, where small inhomogeneities 
grew into a periodically oscillating pattern. Open circles indicate regions
of linear stability where small inhomogeneities decay into a uniform and
static pattern.}
\label{stability}
\end{figure}

If the MinD concentration is increased or decreased beyond 
the limits shown in Fig.~\ref{period_density} 
(left), then the oscillation amplitude
decays, and a  
uniform steady--state results. The stability is mapped out in 
Fig.~\ref{stability}, 
as a function of protein concentration.  This is consistent with 
experiment, where overexpression of MinD suppresses division 
\cite{deBoer1}. Although varying the MinE concentration does effect
the region of oscillatory instability (as shown in
Fig.~\ref{stability}), it did not
have a significant effect on the oscillation period. This appears
somewhat contrary to the results of Ref.~\cite{Raskin}, possibly
due to the absence of MinE dimerization in our model \cite{King}. 

\vspace*{-3mm}

In conclusion, we have introduced a particle--conserving reaction--diffusion
model that self--organizes to form a key regulatory mechanism for
accurate midcell division site selection in {\it E. coli}.  The model 
qualitatively agrees with many of the features found in experiments, and, 
in particular, naturally accounts for the oscillatory patterns 
of the {\it min} proteins.
Already our model leads us to make a number of striking predictions: 
we require that cytoplasmic MinD recruits MinE to the 
membrane; we require that the membrane--associated
diffusion constants for MinD and MinE are very much less than their
corresponding values in the cytoplasm; and we have mapped out the
shape of the oscillation regime as a function of average MinD and MinE
concentration. 

Experimental characterization of reaction rates and
diffusion constants do not yet severely constrain our model. Accurate
experimental measurements of oscillation periods and wavelengths as a
function of concentrations of MinD and MinE will provide a
stringent test.  There is also considerable scope for extending our 
results.  In subsequent studies we will explore a bulk 
$3d$ system with discrete particle dynamics and microscopic
interactions between individual protein molecules. 
This will allow us to explicitly consider the influence of 
fluctuations due to discrete particles, 
the role of ATPase activity of MinD \cite{deBoer}, and
the effects of MinE dimerization \cite{King}.   

This work was supported financially by NSERC Canada. 
We would like to thank Russell Bishop for encouragement and useful
comments and Michael Greenwood for his development of imaging
software. 


\end{document}